# Study of n-n correlations in d + $^2$H → p + p + n + n reaction

**E Konobeevsky[1], A Kasparov[1], M Mordovskoy[1], S Zuyev[1], V Lebedev[2] and A Spassky[2]**

[1] Institute for Nuclear Research, Russian Academy of Sciences, 60-letiya Oktyabrya prospekt 7a, Moscow, 117312, Russia
[2] Skobel'tsyn Institute of Nuclear Physics, Moscow State University, Moscow, 119991, Russia

E-mail: konobeev@inr.ru

**Abstract.** A kinematically complete measurement of the four-body breakup reaction d+$^2$H→$^2$p$^S$+$^2$n$^S$→p +p +n +n has been performed at 15 MeV deuteron beam of the SINP MSU. The two protons and neutron were detected at angles close to those of emission of $^2$p$^S$ and $^2$n$^S$ systems. The energy of singlet dineutron state was determined by comparing experimental TOF spectrum of breakup neutrons with simulated spectra depending on this energy. A low value $E_{nn}$ = 0.076 ± 0.006 keV obtained by fitting procedure apparently indicates an effective enhancement of nn-interaction in the intermediate state of studied reaction.

## 1. Introduction
The goal of this work is experimental study of neutron-neutron interaction and, in particular, of mechanisms of nn-correlations in few-nucleon systems. According to [1] strong discrepancies of experiment and theory observed in nd- and pd-breakup reactions [2, 3] can be explained by a significant strengthening of nn- and pp-correlations of attractive character in the third nucleon field in $^3$H (pnn) and $^3$He (ppn) systems.

In the present work the first results of our study of d + $^2$H → (pp) + (nn) reaction are presented. In this reaction, the correlated nn- and pp-pair can be formed dynamically in the intermediate state. Thus, the measured nn- and pp-correlations, in particular, energies of nn (pp) singlet quasibound states may appear different than those inherent for a free nn- and pp-systems.

## 2. Simulation of the d + $^2$H → p + p + n + n reaction and experimental conditions
The reaction d + $^2$H → (nn)$^s$ + (pp)$^s$ → n + n + p + p is unique as it can proceed through the formation of two-nucleon singlet systems. Kinematic simulation of "quasi-binary" reaction d + $^2$H → $^2$n$^s$ + $^2$p$^s$ allowed one to determine the angles of particle detection in the experiment. It is assumed that the both protons will be detected by the same detector and the neutron will be detected at an angle close to the angle of emission of dineutron system. Further simulation was performed using the kinematical program of reactions with four particles in the final state [4] under the following conditions: $E_d$ = 15 MeV, $\Theta_n$ = 36° ± 1.5°, $\Theta_{p1}$ = $\Theta_{p2}$ = 27° ± 1.5°. The data on the proton energies were used for calculating losses of two protons passing through the same telescope. These calculations show that the area of p+p-events can be unambiguously separated in the experiment from the loci of single particles (p, d, $^3$He).

For each simulated events one can determine the relative energy of two neutrons, i.e., the excess energy of two-neutron system over the two-neutron mass:

$$E_{nn} = [E_1 + E_2 - 2(E_1 \cdot E_2)^{1/2} \cos\Delta\Theta]/2 \qquad (1)$$

Selecting events having excitation energy in the region $E_{nn} \pm \Gamma_{nn}$, where $\Gamma_{nn}$ is a width of selection interval, we would analyze various distributions of output kinematic variables. Earlier [5, 6], we have shown that in reactions with formation and breakup of quasibound NN-states and under certain kinematical conditions one can obtain a specific timing spectrum of breakup particles, characterized by two peaks with a distance between them, depending on the excitation energy $E_{nn}$. The presence of two peaks in the spectrum is due to the fact that in reactions with formation and breakup of quasi-bound state, and provided that the detection of particles occurs at the angle corresponding to emission of NN-system in a two-particle reaction, to hit detector may only breakup particles emitted in c.m. system or in forward (~ 0°) or in backward (~ 180°) direction.

In general, it can be noted that the shape of the timing spectrum is sensitive to both $E_{nn}$ and $\Gamma_{nn}$ values that will allow us to determine these quantities from a comparison of experimental data and simulation results.

### 3. Experimental setup and results

The experiment was performed at a 15 MeV deuteron beam of SINP MSU. In the measurement the $CD_2$-target with thickness of 2 mg/cm$^2$ was used. Two protons were detected by a $\Delta E$-$E$ telescope at the angle of 27º while the neutron was detected at 36º with time-of-flight distance of 0.79 m. In the experiment we used a data acquisition system based on CAEN digitizers DT5742 and DT5720 described in [7]. To calibrate the TOF spectra the neutron detector was transferred to 83º angle for registration of the neutron from two-particle reaction d + $^2$H → $^3$He + n.

Figure 1 represents a two-dimensional $\Delta E$-E plot, obtained at condition of coincident signals from $\Delta E$-, E- and n-detectors, with visible $^3$He, $^4$He, proton and deuteron loci. It is presented also the area for simulated two-proton events. One can see that considerable number of experimental events falls in this area.

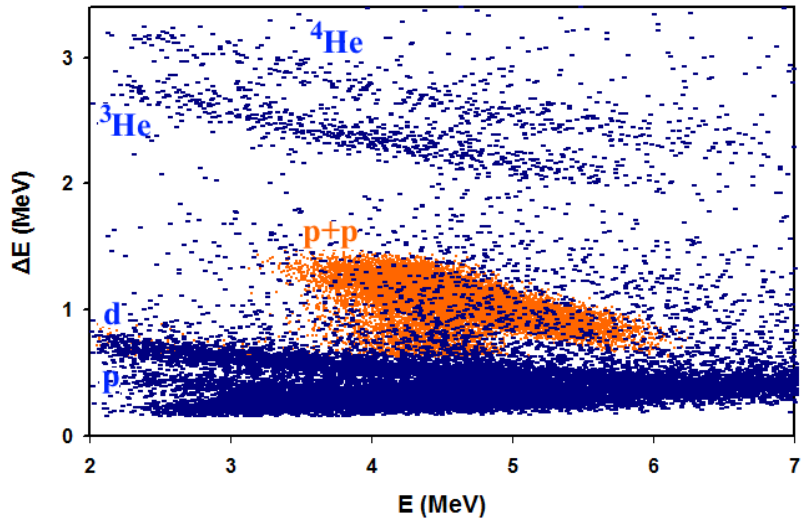

**Figure 1.** Two-dimensional $\Delta E$-E diagram. Dots – experimental events, red area – the simulation results for two-proton passing through the $\Delta E$-E telescope.

The selection of events in this area allow us to obtain the corresponding timing spectrum of neutrons which was compared with the simulation results for various values of the excitation energy $E_{nn}$ and width of the interval $\Gamma_{nn}$. In the process of fitting for each value of $E_{nn}$ and $\Gamma_{nn}$ we calculated $\chi^2$ value for the experimental and simulated spectrum, given by the expression:

$$\chi^2(E_{nn}, \Gamma_{nn}) = \sum_t \frac{(N^{th}_{E_{nn},\Gamma_{nn}}(t) - A \cdot N^{exp}(t))^2}{(\Delta N^{exp}(t))^2}, \qquad (2)$$

where $A$ is a normalization factor, defined as the ratio of the integrals of the experimental and theoretical spectra over the entire interval of summation $t$, and $N^{exp}(t)$ is the statistical error of the experimental points. For data submitted the time summation interval was 31 – 48 ns and included most part of the observed structure. For each fixed value $E_{nn}$ the optimum value of width $\Gamma^{opt}_{nn}$ was selected by a minimum value of $\chi^2$ (figure 2). Thus, the minimum value has been found for each value $E^i_{nn}$.

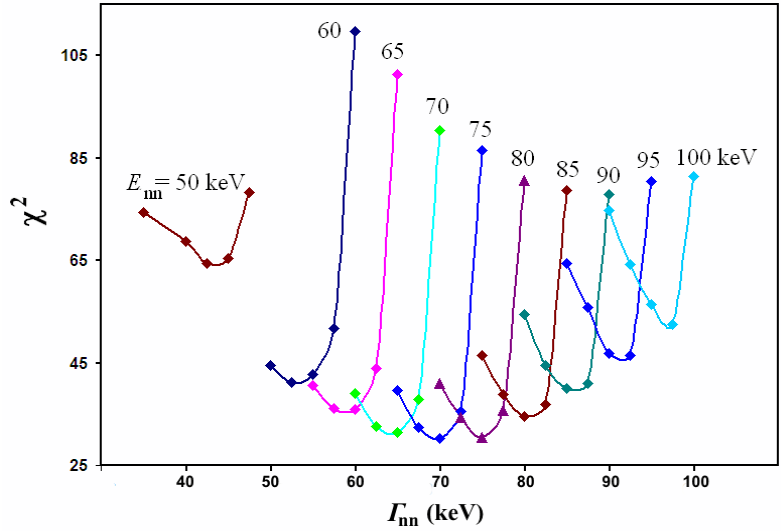

**Figure 2.** Dependence of $\chi^2$ values on $\Gamma_{nn}$ for various $E^i_{nn}$.

In figure 3 the experimental spectrum is compared with various variants of simulated spectra: for democratic breakup ($E_{nn} = 0 – 1300$ keV), $E_{nn} = 160$ keV, $\Gamma_{nn} = 150$ keV; $E_{nn} = 80$ keV, $\Gamma_{nn} = 70$ keV and $E_{nn} = 40$ keV, $\Gamma_{nn} = 35$ keV. For the latter three variants, simulation for the indicated $E^i_{nn}$ values was performed with optimal value of the parameter $\Gamma^{opt}_{nn}$.

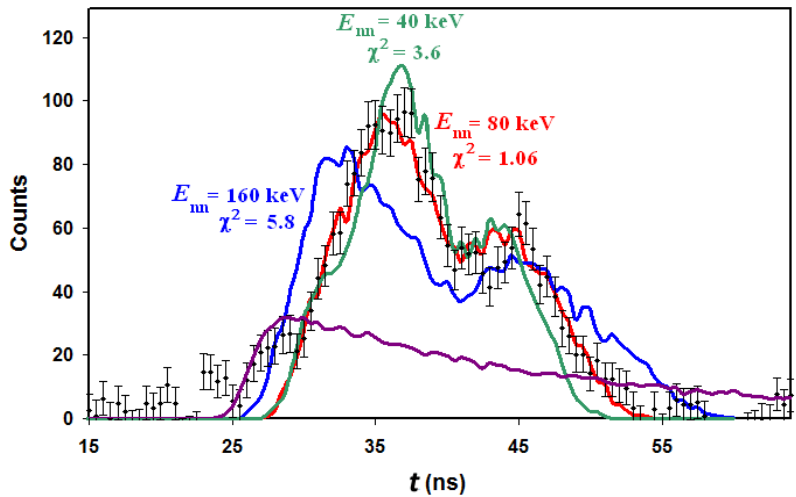

**Figure 3.** Experimental and simulated neutron TOF spectra.

To determine $E_{nn}$ the dependence of $\chi^2(E_{nn})$ value on $E_{nn}$ was fitted by a quadratic polynomial (figure 4). Thus, the minimum value of $\chi^2$ determines the most probable value of the quasibound state energy, and error in determining $E_{nn}$ is given as

$$\Delta E_{nn} = \left| E_{nn}(\chi^2_{min}) - E_{nn}(\chi^2_{min} + 1) \right| \qquad (3)$$

The minimum value of the polynomial is achieved at $E_{nn}$ = 76 keV, $\Delta E_{nn}$ = ± 6 keV.

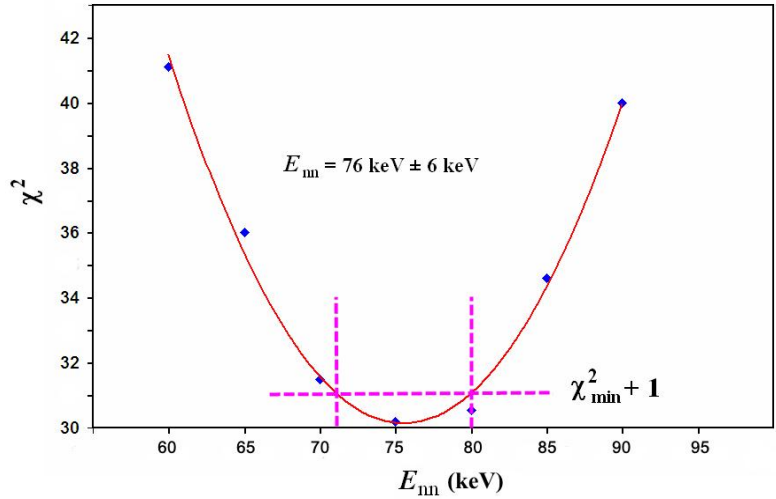

**Figure 4.** Dependence of $\chi^2(E^i_{nn}, \Gamma^{opt}_{nn})$ on the value of $E^i_{nn}$. The curve is a result of fitting by a quadratic polynomial.

## 4. Conclusions
We investigated the d + $^2$H → $^2$n$^s$ + $^2$p$^s$ → n + n + p + p breakup reaction, passing through a formation in the intermediate state of dineutron and diproton singlet pairs. For the first time, in a kinematically complete experiment the energy of quasi-bound state of $^2$n-system is determined. This value is determined by comparing experimental TOF spectrum of neutrons from breakup of this state with simulated spectra depending on this energy. The obtained value $E_{nn}$ = 76 ± 6 keV is significantly lower than energies ($E_{nn}$ = 120 – 160 keV) recalculated from the experimental values of $^1S_0$ nn-scattering length [8]–[11] that apparently indicates an effective enhancement of nn-interaction in the intermediate state of studied reaction.